\documentclass[twocolumn,prl]{revtex4}
\usepackage{amssymb}
\usepackage{graphicx}
\usepackage{natbib}
\usepackage{amsmath}

\begin{document}

\title{Universal temperature dependence of optical excitation life-time and band-gap in chirality assigned semiconducting single-wall carbon nanotubes}

\author{Ferenc Simon$^{\dag}$}
\author{Rudolf Pfeiffer}
\author{Hans Kuzmany}
\affiliation{Institut f\"{u}r Materialphysik, Universit\"{a}t Wien, Strudlhofgasse 4,
A-1090 Wien, Austria}

\begin{abstract}

The temperature dependence of optical excitation life-time,
$1/\Gamma$, and transition energies, $E_{\text{ii}}$, were measured
for bucky-papers of single-wall carbon nanotubes (SWCNTs) and inner
tubes in double-wall carbon nanotubes (DWCNTs) using resonant Raman
scattering. The temperature dependence of $\Gamma$\ and
$E_{\text{ii}}$ is the same for both types of samples and is
independent of tube chirality. The data proves that electron-phonon
interaction is responsible for temperature dependence of
$E_{\text{ii}}(T)$. The temperature independent inhomogeneous
contribution to $\Gamma $ is much larger in the SWCNT samples, which
is explained by the different SWCNT environment in the two types of
samples. $\Gamma$ of the inner tubes for the bucky-paper DWCNT
sample is as low as $\sim$ 30 meV, which is comparable to that found
for individual SWCNTs.
\end{abstract}

\maketitle


Single-wall carbon nanotubes (SWCNTs) have been in the focus of
interest in the last decade due to their unique physical and
chemical properties, which
makes them a potential candidate for a broad range of applications \cite%
{DresselhausTubes,DresselhausTubesNew}. It is now established theoretically \cite%
{KanePRL2003,LouiePRL2004,AvourisPRL2004,AvourisPRL2005} and there
is emerging
experimental evidence \cite%
{HeinzSCI2005,MaultzschPRB2005,JorioPreprint,MaruyamaPreprint} that
excitonic effects, i.e. the correlation of photoexcited electrons
and holes, play an important role in the optical absorption and
emission properties of single-wall carbon nanotubes (SWCNTs). The
life-time of the excitons, $\tau_\text{exciton} $, and the optical
excitation energies, $E_{\text{ii}}$, and their temperature
dependencies are crucial parameters for the optoelectronic
applications of SWCNTs. SWCNTs are characterized by the great
variation of physical properties depending on the ($n,m$) chiral
indices. Therefore chirality assigned determination of these
parameters is vital. $\tau_\text{exciton}$ was measured in
time-resolved photoexcited studies
\cite{LauretPRL2003,HeinzPRL2004,SmalleyPRL2004,HartschuhPRL2005}
and estimated theoretically \cite{LouiePRL2005,AvourisNL2005}.
E$_{\text{ii}}$ was measured using photoluminescence spectroscopy in
a chirality assigned manner for ensembles of isolated SWCNTs
\cite{Bachilo:Science298:2361:(2002)} and for individual SWCNTs
\cite{LefebvrePRL2003,HartschuhSCI2003}.

The scattering rate of the optically excited states, $\Gamma=1/\tau
$, is related to $\tau_\text{exciton}$. Both $\Gamma$ and
E$_{\text{ii}}$ are accessible from resonant Raman scattering (RRS)
studies \cite{Falicov,KuzmanyEPJB}. Such studies have been presented
on surfactant separated \cite{FantiniPRL2004,TelgPRL2004} and also
for bucky-paper SWCNT samples \cite{FantiniPRL2004}. The temperature
dependence, studied using varying laser powers, of $E_{\text{ii}}$
was found to markedly differ for $\nu=1$ and $\nu=2$ type
semiconducting SWCNTs, where $\nu$ is given by $(n-m)\mod{3}=\nu$.
$\nu=1$ tubes were found to blue-shift whereas $\nu=2$ to red-shift
with increasing temperature. In a recent study, E$_{\text{ii}}$ was
found to red-shift for both $\nu=1,2$ with increasing temperature
for individual SWCNTs \cite{CapazPRL2006}, which was explained by
the electron-phonon coupling \cite{CapazPRL2005,CapazPRL2006}. The
anomalous dependence observed for bucky-paper samples in Ref.
\cite{FantiniPRL2004} is still unresolved. Clearly, the role for the
tube environment requires further studies. In addition, no
temperature dependent measurement of $\Gamma$ on chirality assigned
SWCNTs is available.

Here, the temperature dependence of the $(n,m)$ specific electronic
transitions and scattering rates for optical excitations is reported
for SWCNTs in different environment. We measured SWCNTs with the
same chirality in bucky-paper SWCNT samples and as inner tubes in
double-wall carbon nanotube (DWCNT) samples. The latter type of
sample has attracted considerable interest as the inner tubes are in
a well-shielded environment, which results in exceptionally long
phonon life-times \cite{PfeifferPRL2003,SimonCPL2005}. The current
study explains the role of the tube environment on the two
parameters studied. A carbon nanotube is surrounded by other tubes
with random chiralities in an SWCNT sample, however a given
inner-tube can be embedded in outer tubes with a few well defined
chiralities. In addition, the narrow Raman line-widths of the
inner-tubes result in distinct inner-tube Raman spectra of the
studied tube-pairs \cite{PfeifferPRL2003,PfeifferPRB2005}. We indeed
observe a large inhomogeneous contribution to $\Gamma$ for the SWCNT
sample but the same temperature dependence as for the inner tubes in
the DWCNT. The same temperature dependence of $E_{\text{ii}}$ was
observed for both types of samples. These results suggest that both
parameters are independent of the tube environment and suggest a
common and robust mechanism for them.


The SWCNTs we studied were HiPco (Carbon Nanotechnologies Inc.
Houston, USA)\ and CoMoCat samples (SouthWest NanoTechnologies Inc.
Oklahoma, USA) with mean tube diameters of $d=$ 1.0 nm and $d=$ 0.8
nm, respectively as determined from a Raman analysis
\cite{KuzmanyEPJB}. The DWCNTs were produced from SWCNTs
encapsulating fullerenes, so-called peapods \cite{SmithNAT}. The
host SWCNTs (Nanocarblab, Moscow, Russia) with a mean diameter of
1.40 nm were sealed under vacuum in a quartz tube together with the
C$_{60}$ powder and heated at 650 $^{\circ}$C for 2 hours
\cite{KatauraSM2001}. The resulting peapods were annealed in dynamic
vacuum at 1270 $^{\circ}$C for 2 hours which results in DWCNTs
\cite{BandowCPL2001,PfeifferPRL2003}. The inner tube diameter
distribution follows that of the outer tube \cite%
{AbePRB2003,SimonPRB2005} and we obtained $d=$ 0.7 nm for the mean
diameter from a Raman analysis \cite{SimonPRB2005}. As a result, the
diameter distribution of SWCNTs and inner tubes are very similar for
the three samples.

Raman spectroscopy was performed on a Dilor xy triple grating
instrument equipped with a home-built cryostat for the 80-600 K
temperature range. The excitation was provided by a tunable laser in
the 2.01-2.18 eV (616-568 nm) energy range involving 22 different
laser energies. A series of discharge calibration lamps and Si
powder were used to calibrate the Raman shifts and the Raman
intensities, respectively. The samples were mounted on the cold
finger of a liquid nitrogen cooled cryostat with conducting silver
for good thermal contact. The spectrometer was operated in the
"macro" acquisition mode with typical laser power densities of 5
mW/(100 $\mu $m)$^{2}$ to avoid sample heating.

\begin{figure}[tbp]
\includegraphics[width=1.0\hsize]{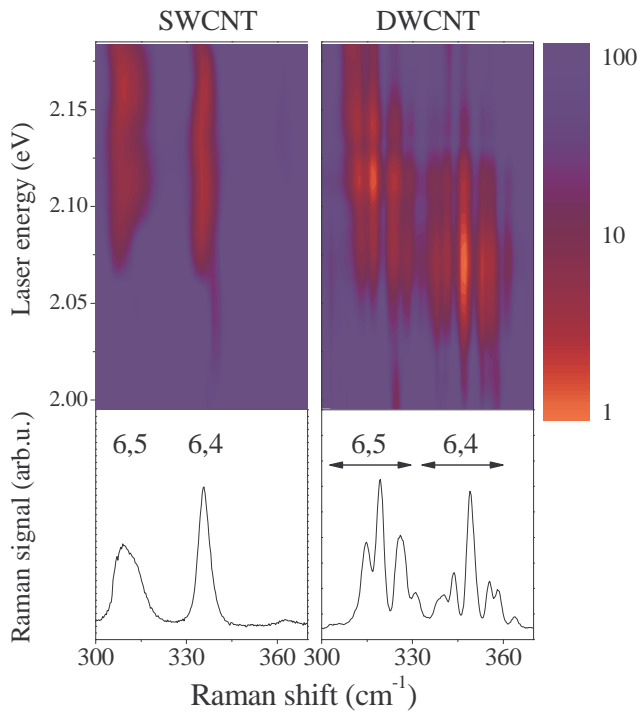}
\caption{\textit{Color online} Energy dispersive Raman contour plot
for SWCNT(CoMoCat) and DWCNT samples at 80 K. The vertical bar shows
the color coding for the observed intensities on a logarithmic scale
in arbitrary units. Lower panels show individual Raman spectra with
2.10 eV laser energy.} \label{ErgDisp}
\end{figure}

In Fig. \ref{ErgDisp} we show the energy dispersive Raman map for
the SWCNT(CoMoCat) and DWCNT samples for a part of the radial
breathing mode (RBM) and excitation energy range at 80 K. Similar
contour plots showing a larger area, have been used
to assign $(n,m)$\ indexes to the different nanotubes in SWCNT \cite%
{FantiniPRL2004,TelgPRL2004}\ and DWCNT samples
\cite{PfeifferPRB2005b}. The two peaks at 310 and 337 cm$^{-1}$ with
2.12 and 2.08 eV transition energies correspond to the $E_{22}$
optical transition enhanced RBMs of the (6,5) and (6,4) tubes for
the SWCNT sample \cite{FantiniPRL2004,TelgPRL2004}. These two tubes
are representative for both semiconducting SWCNT classes $\nu=1$ and
$\nu=2$ and their RBMs are well separated from all other tube modes.
This renders them ideal for studying the temperature dependent RRS.
For the DWCNT, (6,5) and (6,4) are inner tubes and their RBMs are
split into several
components as the same inner tube can be in several outer ones \cite%
{PfeifferEPJB2004,PfeifferPRB2005b}. The difference in the resonance
energies results from the different environment for the two
samples:\ an outer tube for the DWCNT sample and the surrounding
ensemble of the other SWCNTs for the SWCNT sample. The gradual
red-shift for the inner tube resonance energies with increasing
Raman shift was associated to the pressure induced by the outer
tubes \cite{PfeifferPRB2005b}.

\begin{figure}[tbp]
\includegraphics[width=1.0\hsize]{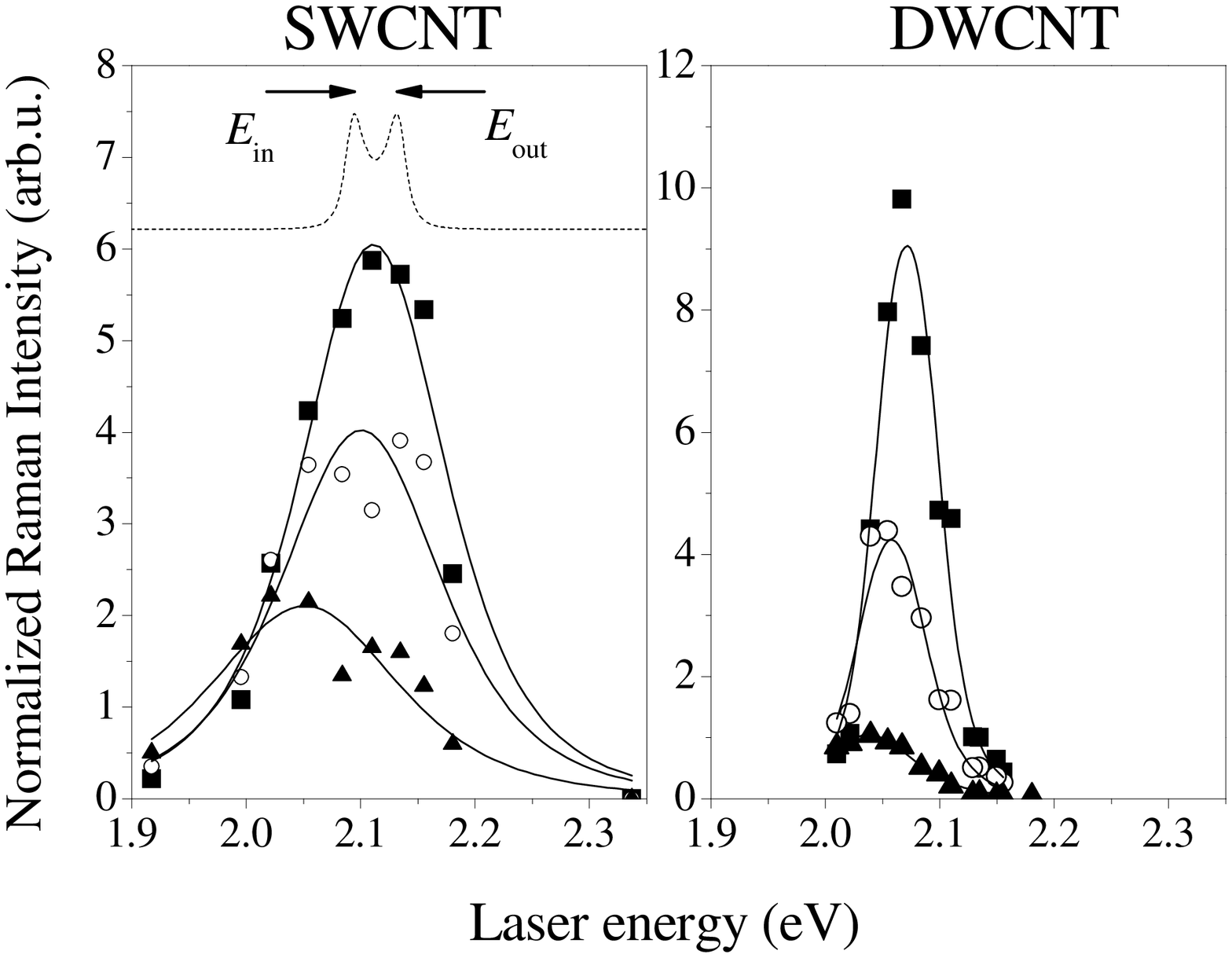}
\caption{Raman resonance profile for the (6,4) tubes in the SWCNT
(CoMoCat) and DWCNT samples, $\blacksquare$: 80 K, $\bigcirc$: 300
K, $\blacktriangle$: 600 K. Solid curves show fits with the RRS
theory. Dashed curve is a simulation for the 80 K SWCNT data with
$\Gamma=10$ meV. Arrows indicate the incoming and outgoing resonance
energies. Note the much narrower widths for the DWCNT sample.}
\label{T_dep_Erg_Disp}
\end{figure}

The temperature dependence of the Raman contour plots was measured
for all three samples. Raman intensities corresponding to individual
modes were determined by fitting Voigtian line-shapes whose Gaussian
component was given by the residual spectrometer resolution. The
energy dependent Raman intensities for two particular tube modes in
the two kinds of samples are shown in Fig. \ref{T_dep_Erg_Disp} as a
function of temperature. For the SWCNT CoMoCat sample the (6,4) tube
mode at 337 cm$^{-1}$ (phonon energy $E_{\text{ph}}=41.8$ meV) is
shown and the strongest (6,4) inner tube component at 347 cm$^{-1}$
($E_{\text{ph}}=43$ meV) is shown for the DWCNT sample. The
temperature dependent resonant Raman data can be fitted with the
resonance Raman theory for Stokes Raman modes
\cite{Falicov,KuzmanyEPJB}:
\begin{eqnarray}
I(E_{\text{l}})  =M_{\text{eff}}^{4}\left\vert
\frac{\left(E_{\text{l}}-E_{\text{ph}}\right)^4\left(n_\text{BE}(E_{\text{ph}})+1\right)}{\left(
E_{\text{l}}-E_{\text{ii}}-i\Gamma \right) \left(
E_{\text{l}}-E_{\text{ph}}-E_{\text{ii}}-i\Gamma \right)
}\right\vert ^{2} \label{Resonance_Raman}
\end{eqnarray}

\noindent Here, the electronic density of states of SWCNTs is
assumed to be a Dirac function and the effective matrix element,
$M_{\text{eff}}$, describing the electron-phonon interactions is
taken to be independent of temperature and energy. $E_{\text{l}}$,
$E_{22}$ and $E_{\text{ph}}$ are the exciting laser, the optical
transition and the phonon energies, respectively.
$n_\text{BE}(E_{\text{ph}})=(\exp(
E_{\text{ph}}/\text{k}_{\text{B}}T)-1)^{-1}$ is the Bose-Einstein
function and accounts for the thermal population of the vibrational
state \cite{KuzmanyBook} and $n_\text{BE}(E_{\text{ph}})+1$ changes
a factor $\sim$ 2 between 80 and 600 K. The temperature dependence
of $E_{\text{ph}}$ is $\sim$ 1 \% for the studied temperature range
\cite{AjayanPRB2002} thus it can be neglected. The first and second
terms in the denominator of Eq. \ref{Resonance_Raman} describe the
incoming and outgoing resonances, respectively and are indicated on
a simulated curve by arrows in Fig. \ref{T_dep_Erg_Disp}. These are
separated by $E_{\text{ph}}$. This means the apparent width of the
resonance Raman data does not represent $\Gamma$.

 Solid curves in Fig.
\ref{T_dep_Erg_Disp} show fits for the data at all temperatures
simultaneously using Eq. \ref{Resonance_Raman} and
 allow thus to derive the temperature dependence for
 $E_{22}$ and $\Gamma$. The simultaneous fit improves its
 reliability compared to fitting separately to the particular temperatures such
 as in Ref. \cite{CapazPRL2006}. The result for the temperature dependent relative optical transition energies, $%
E_{22}(T)-E_{22}(80\text{ K})$ and $\Gamma (T)$ is summarized in Fig. \ref%
{T_dep_params}. The room temperature values of $E_{22}$ is given in Table %
\ref{Table} together with the $\Gamma (80\text{ K})$ and the $\Delta
\Gamma =\Gamma (600K)-\Gamma (80K)$ parameters for the three samples
studied.

\begin{figure}[tbp]
\includegraphics[width=0.8\hsize]{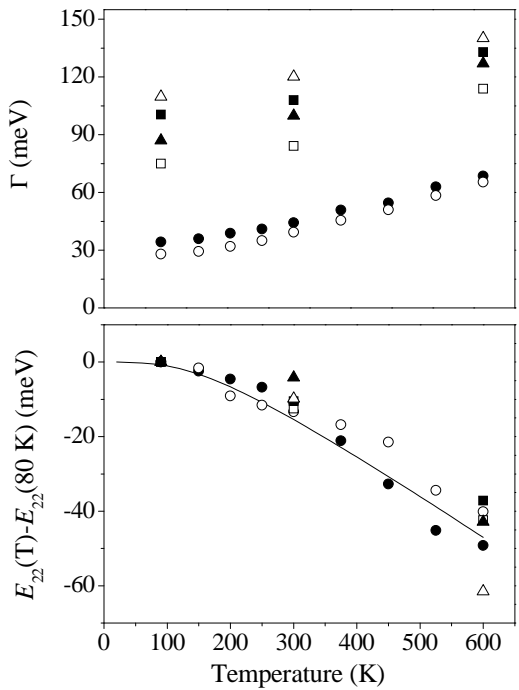}
\caption{Temperature dependence of the optical transition energy
shift, $E_{22}(T)-E_{22}(80\text{ K})$, and the damping parameter,
$\Gamma $ for the $(6,5)$ and $(6,4)$ inner tubes of DWCNTs, HiPco
tubes, and CoMoCat tubes. Open and closed symbols are for the (6,4)
and (6,5) tubes, respectively. $\triangle$: CoMoCat, $\Box$: HiPco,
$\bigcirc$: DWCNT. Solid curve is a calculation for the red-shifting
$E_{22}$ for a (6,5) tube as explained in the text. }
\label{T_dep_params}
\par
\end{figure}

\begin{table}[tbp]
\caption{Transition energies, E$_{22}$, and the damping parameter,
$\Gamma$ in the two SWCNT and the DWCNT samples. The HiPco sample
dispersed in SDS were studied at room temperature in Ref.
\cite{FantiniPRL2004}. Relative errors are $2 \cdot 10^{-3}$ for
$E_{22}$ and $0.1$ for $\Gamma$.}

\label{Table}
\begin{tabular}{lllllll}

\hline \hline & \multicolumn{2}{c}{E$_{22}$ (eV)} &
\multicolumn{2}{c}{$\Gamma(80\text{ K})$ (meV)}
& \multicolumn{2}{c}{$\Delta\Gamma$ (meV)} \\
\hline
& $(6,5)$ & $(6,4)$ & $(6,5)$ & $(6,4)$ & $(6,5)$ & $(6,4)$ \\
CoMoCat & 2.12 & 2.08 & 90 & 109 & 40 & 31\\
HiPco & 2.11 & 2.07 & 100 & 77 & 33 & 38 \\
HiPco SDS \cite{FantiniPRL2004,Fantinicomment} & 2.18 & 2.11 & 35 &
35 & -- & -- \\
DWCNT & 2.09 & 2.04 & 34 & 27 & 34 & 38\\

\hline \hline
\end{tabular}
\end{table}

We start the discussion with the $T$ dependent optical transition
energies. Differences between the room temperature values for
E$_{22}$'s for DWCNT \cite{PfeifferPRB2005b}, SWCNT in bucky-paper
and as dispersed \cite{FantiniPRL2004} have been previously observed
and explained by the effect of the different environment. The
surprising observation is the overall red-shift of 50(10) meV
between 80 and 600 K irrespective of the sample type and the family
of tube, $\nu=1$ or 2. Recently, $\sim$ 50 meV red-shifts were
observed for individual suspended nanotubes using the same method
for SWCNTs with $\nu_{\text{RBM}}>250 \text{cm}^{-1}$
\cite{CapazPRL2006}. The observed temperature dependence was
explained by the softening of the band gap at high temperatures due
to electron-phonon coupling \cite{CapazPRL2005,CapazPRL2006}
yielding the optical transition energy shift:

\begin{eqnarray}
\Delta
E_\text{22}(T)=\alpha_{1}\Theta_{1}n_{\text{BE}}(\text{k}_{\text{B}}\Theta_{1})+\alpha_{2}\Theta_{2}n_{\text{BE}}(\text{k}_{\text{B}}\Theta_{2})
\label{Capaz_curve}
\end{eqnarray}

In Fig. \ref{T_dep_params}, we show the calculated $\Delta
E_\text{22}(T)$ using Eq. \ref{Capaz_curve} with
$\alpha_{1}=-1.17\cdot 10^{-5}$ eV/K, $\alpha_{2}=-1.04 \cdot
10^{-4} $ eV/K, $\Theta_{1}=103.8$ K, and $\Theta_{2}=487.4$ K for a
(6,5) tube from Ref. \cite{CapazPRL2006}. Since $\Delta
E_\text{22}(80\text{ K})\approx \Delta E_\text{22}(0 \text{ K})$
plotting $\Delta E_\text{22}(T)$ on the $E_{22}(T)-E_{22}(80\text{
K})$ data is justified. A similar calculation for a (6,4) tube does
not give a significantly different curve. A good agreement between
the experimental points and the calculation is observed, which
confirms the validity of the theoretical explanation presented in
Ref. \cite{CapazPRL2005}. Both ours and the results in Ref.
\cite{CapazPRL2006} are in difference to a previous study in which a
blue-shift with increasing temperature was reported for $\nu=1$
tubes in a laser heated sample \cite{FantiniPRL2004}. Since the
current experimental conditions provide a reliable control over the
true sample temperature the samples studied here behave differently
as compared to those studied previously.

Our result confirms the red-shift of optical transition energies
with increasing temperature as observed on individual tubes
\cite{CapazPRL2006}. It holds for SWCNTs in bucky-papers and for
inner tubes surrounded by outer tubes in DWCNT samples. It thus
unambiguously proves that electron-phonon interaction is the
dominant mechanism for the temperature variation of the optical
transitions and other effects such as thermal expansion, proposed in
Ref. \cite{CapazPRL2006} do not play a role.

The temperature dependent $\Gamma(T)$ can be separated into a
residual and a $T$ dependent part. The measurement for the lowest
temperatures, the $\Gamma(\text{80 K})$ values, approximate the
residual $\Gamma_{0}$ which shows marked differences for the three
samples. In contrast the change with temperature, $\Delta \Gamma
=\Gamma (\text{600 K})-\Gamma (\text{80 K})$, is similar for all
samples and chiralities studied and scatters in the 20-30 meV range.
We assign the temperature dependent part of $\Gamma(T)$ to
homogeneous thus to a true life-time effect and the residual
$\Gamma_{0}$ to inhomogeneous broadening.

The low value of $\Gamma_{0}$ for individual SDS dispersed SWCNTs
and inner tubes in DWCNTs and its high value for the two bucky-paper
samples of SWCNTs can be explained by the sample morphologies
assuming that nearest neighbor tube-tube interactions strongly
affect the transition energies. For the bucky-paper SWCNT samples,
each tube is surrounded by other tubes with random chiralities. The
interaction between the tubes gives rise to a distribution of
transition energies, which appears as an inhomogeneous broadening of
the RRS. For individual SWCNTs the environment is the same for all
tubes, thus transition energies are well defined. For inner tubes in
the DWCNT bucky-paper sample, an inner tube can be embedded in outer
tubes with several possible chiralities
\cite{PfeifferEPJB2004,PfeifferPRB2005b}. However these inner-outer
tube pairs are well resolved spectroscopically for most of the inner
tubes and in particular for the (6,5) and (6,4) tubes studied. Thus
the inner tube transition energies are well defined for each
inner-outer tube pair and $\Gamma$ is not affected by inhomogeneous
effects.

The current data for $\Gamma$ also account for the observation of
the extremely large Raman cross section of inner tubes as compared
to SWCNTs \cite{PfeifferPRL2003,PfeifferPRB2005,SimonPRL2005}. The
factor two smaller $\Gamma$ of inner tubes results in an approximate
enhancement of a factor 16 in the Raman cross section as the latter
is proportional to
$\left({\Gamma^{2}\left(E_{\text{ph}}^{2}+\Gamma^{2}\right)}\right)^{-1}$.

In conclusion, we studied the temperature dependent resonant Raman
scattering on small diameter SWCNTs and inner tubes in DWCNTs. An
overall red-shift of the optical transition energies with the same
magnitude irrespective of tube type was observed. We observed the
same $T$ dependence of the inverse optical excitation life-time,
$\Gamma$, which is attributed to fundamental on-the tube scattering
mechanisms. The environment gives rise to an inhomogeneous
broadening of the RRS resonance profiles and results in a small
$\Gamma$ for the inner tubes. Interestingly, $\Gamma$ for inner
tubes in a bucky-paper sample of DWCNTs is identical for a
micelle-suspended SWCNTs. Thus the current result underlines the
application potential of DWCNTs in future optoelectronic devices.
The measurement of the temperature dependence of $\Gamma$ and of the
optical transition energies provides important input for theories
explaining the optical properties of carbon nanotubes.

Supported by the FWF project Nr. 17345 and by the EU Grants
MERG-CT-2005-022103 and BIN2-2001-00580. F.S. acknowledges the
Zolt\'{a}n Magyary postdoctoral programme and the Hungarian State
Grants (OTKA) No. TS049881, F61733 and NK60984 for support.

\dag corresponding author: ferenc.simon@univie.ac.at \\ Present
address: Budapest University of Technology and Economics, Institute
of Physics and Solids in Magnetic Fields Research Group of the
Hungarian Academy of Sciences, H-1521, Budapest P.O.Box 91, Hungary

\bibliographystyle{apsrev}


\end{document}